\documentclass{article}
\usepackage{graphicx} % Required for inserting images

\title{Text2SQL}
\author{gladtv }
\date{March 2025}

\begin{document}

\maketitle

\section{Introduction}

Natural language interfaces for databases have become increasingly important as organizations seek to democratize data access and enable non-technical users to extract insights through simple questions. Text2SQL systems address this need by translating natural language queries into structured SQL code that can be executed against a database. However, developing effective Text2SQL systems presents significant challenges, mainly when dealing with complex database schemas, domain-specific terminology, and the nuanced ways users express their information needs.

This document outlines a comprehensive approach to building a robust Text2SQL system using a Retrieval-Augmented Generation (RAG) framework, which incorporates domain knowledge, database structure, and example-based learning to enhance accuracy and reliability. Our system's core is a knowledge base built from database documentation and successful query examples. We store this information in a vector database and use retrieval techniques to find the most relevant context for each user question, helping us generate accurate SQL.

The described approach tackles common problems in SQL generation, including selecting the wrong tables, creating incorrect joins, and misunderstanding business rules. By following a straightforward process of knowledge collection, context retrieval, and SQL generation, we managed to produce reliable results while adapting to different types of questions and database structures.

The following sections explain how we train the knowledge base, retrieve relevant information, and generate SQL code, providing a clear guide for implementing this solution. We provide a detailed description of the methodology for training the knowledge base, retrieving relevant context, and generating SQL code, offering a clear guide for implementing an effective Text2SQL solution in real-world applications.

\section{General Concept}

The Text2SQL system is designed to convert natural language queries into accurate SQL code by using a sophisticated RAG (Retrieval-Augmented Generation) framework. This framework operates based on a comprehensive Knowledge Base that combines multiple sources of information:

1. \textbf{Database Documentation}: Detailed descriptions of database tables, columns, and relationships within the overall system. This provides the basis for understanding the database schema.
2. \textbf{Question-Answer Examples}: A collection of question-SQL pairs that is used as a reference for mapping natural language questions to specific SQL syntax and database operations.
3. \textbf{Domain-Specific Rules}: Specialized instructions that capture the business logic and relationships not explicitly stored in the vector database but critical for accurate query generation.

The system functions through three main phases:

1. \textbf{Knowledge Base Training}: The process of building a comprehensive understanding of the database by analyzing documentation, examples, and schema information. This involves transforming unstructured descriptions into structured metadata and vector embeddings.
2. \textbf{Content Retrieval}: This phase includes retrieving the most relevant documentation and examples from the knowledge base using semantic similarity matching with the user’s request concerning the business rules.
3. \textbf{SQL Generation}: The final stage is where the system utilizes the retrieved information to generate a correct SQL query that satisfies the user's intent, leveraging both table descriptions and similar query examples.

This approach ensures that the Text2SQL system can handle complex queries by combining structural knowledge about the database with practical examples of query patterns, all while respecting domain-specific business rules that govern data relationships and access patterns.

\section{Train Knowledge Base}

The Text2SQL model training consists of two parts: Vector DB training in the textual context, which includes documentation, database schema, DDL, and ground truth examples of correct answers (SQL code) for the user's request and extraction of domain-specific rules. The latter is not stored in the vector DB; instead, it is used as an additional context (together with business rules) to increase the precision of the AI agent response.

\subsection{Train Based on Documentation}

The training of the Text2SQL model's Knowledge Base, which is the core part of the Text2SQL RAG framework, leads to saving the valuable content in Vector DB. The process starts with training on documentation. Each document is represented in JSON format and contains information about a specific table in the database with its detailed description, including information about columns and connections with other datasets in the database. The document should have the following structure:
\begin{itemize}
\item \textbf{name}: The table’s name
\item \textbf{summary}: A brief description of what the table contains
\item \textbf{purpose}: Explains the table’s role and why it exists
\item \textbf{dependencies\_thoughts}: Informal notes on relationships between this table’s fields and fields in other tables
\item \textbf{keys}: A list of key columns used to join or relate to other tables
\item \textbf{connected\_tables}: An array of related table names
\item \textbf{columns}: An array where each element is an object describing a column (with its name and a description)
\item \textbf{entities}: A list of key concepts extracted from the documentation
\item \textbf{strong\_entities}: A subset of entities deemed particularly significant
\end{itemize}

You don't need to manually create this structural definition. The system can automatically generate it by analyzing any free-form text that contains descriptions of the table and its columns. This automated approach saves time and reduces the potential for errors that might occur during manual creation. The system can extract the necessary information about table relationships, column definitions, and data types directly from natural language descriptions, converting them into a formal structure without requiring manual intervention.

\subsection{Documents Processing}

The automated extraction of the required fields from a document is performed via processing the document by LLM and divided into two steps:

1. Extraction of \texttt{name}, \texttt{summary}, table \texttt{purpose}, \texttt{columns} with their description, \texttt{key} columns, and \texttt{connected tables}.

2. Extraction of \texttt{entities} that might be inferred from this table based on its purpose.

For the second step, the voting strategy is applied - we generate several LLM responses for the same input content and assess the frequency of the concept occurrence within multiple runs. This allows extracting the strong entities from the entity list.

\subsection{Documents Auto-Generation}

If the user's DB contains well-named tables with columns that also have explicit names (the names reflect the columns' and table's sense), we may assume their purposes and dependencies considering the domain and the general knowledge about available data. So, if we have a DB schema, we may use LLM to generate the table description with some reasonable limitations. Generally, the generated description is applicable in the first approximation, but domain expert validation is strongly required for complex data structures.

\textbf{Key Characteristics of Generated Data}:

1. Autogenerated descriptions follow a clear format with sections like "Table Description," "Columns Description" and "Potential Dependencies," making them easy to navigate and review.

2. Provide detailed descriptions of columns, often attempting to explain their possible roles and relationships.

3. Highlight primary keys, sort keys, and potential table dependencies, which can be helpful in understanding the database structure and optimizing queries.

\textbf{Limitations}:

1. Autogenerated descriptions may sometimes define the role of a table too broadly rather than focus on its specific function.

2. Descriptions may suggest dependencies between tables that are not explicitly defined, leading to incorrect assumptions.

3. General column descriptions may not always capture specific nuances, such as predefined status codes or intentionally unused fields.
\end{document}

\section*{Entity Extraction Process}

Once all responses are received, the next step is to combine the extracted entities from each response 
into one comprehensive collection. This aggregation allows the system to analyze the frequency 
with which each unique entity appears across the multiple outputs. The underlying assumption is 
that entities mentioned consistently are more likely to be accurate or significant. Those entities 
occur just once within the series of LLM calls and are considered insignificant. This approach 
leverages the diversity of multiple model outputs to enhance the reliability of entity extraction. By 
aggregating and filtering based on frequency, the process reduces the impact of any anomalous or 
less confident responses, resulting in a more robust and consistent set of extracted entities.

\section*{EXAMPLE}

Let us consider an example. Suppose we have well-documented database with the detailed description 
of each table, which looks like:

\begin{Verbatim}[frame=single]
SMS Statuses
------------
The `sms_statuses` table contains data about which activists are subscribed to the mobile messaging list of groups being mirrored (or children, if in a network) and what their current statuses are (subscribed, unsubscribed, bouncing, or spam complaint).
This table contains data about the mobile subscription status such what status the activist has currently, as well as metadata about the mobile subscription status such as when it was created.
#### Fields
| Field Name | Description | Type | Is Sort Key? |
| --- | --- | --- | --- |
| id | The numerical identifier of the mobile subscription status. | INT |  |
| subscriber_id | The numerical identifier of the activist who is attached to this status.  Relates to the `id` field in the `[users](/mirroring/docs/users)` table. | INT | True |
| group_id | The numerical identifier of the group who's list this mobile subscription status is related to.  Relates to the `id` field in the `[groups](/mirroring/docs/groups)` table. | INT |  |
| user_id | This field is intentionally blank. | INT |  |
| status | The mobile subscription status of the activist. `1` if subscribed. `2` if unsubscribed. `3` if bouncing. | INT |  |
| source_action_id | This field is intentionally blank. | INT |  |
| source_action_type | This field is intentionally blank. | VARCHAR |  |
| join_date | The UTC timestamp when this mobile subscription status was last changed from any status (or none) to subscribed. Ex: The date the person last joined the list. | DATETIME |  |
| created_at | The UTC timestamp when this mobile subscription status was created. | DATETIME |  |
| updated_at | The UTC timestamp when this mobile subscription status was last updated. | DATETIME |  |
\end{Verbatim}

It is structured representation, which is completely ready for use, will have a view:

\begin{Verbatim}[frame=single]
{
  "name": "sms_statuses",
  "summary": "Data about the mobile subscription statuses of activists within groups.",
  "purpose": "To track and manage the subscription statuses (subscribed, unsubscribed, bouncing, etc) of activists to mobile messaging lists.",
  "dependencies_thoughts": "Relates to the `users` table (subscriber_id to id) and the `groups` table (group_id to id)",
  "keys": [
    "subscriber_id",
    "group_id"
  ],
  "connected_tables": [
    "users",
    "groups"
  ],
  "columns": [
    {
      "column": "id",
      "description": "The numerical identifier of the mobile subscription status."
    },
    ...
    {
      "column": "updated_at",
      "description": "The UTC timestamp when this mobile subscription status was last updated."
    }
  ]
}
\end{Verbatim}